\documentclass[10pt,letterpaper]{article}
\usepackage{opex3}
\usepackage{epsfig}
\usepackage{amsmath}

\newcommand{\rrr}{\boldsymbol r}
\newcommand{\EEE}{\boldsymbol E}
\newcommand{\DDD}{\boldsymbol D}
\newcommand{\nablabf}{\boldsymbol \nabla}

\begin{document}

\title{Air-clad fibers: pump absorption assisted by chaotic wave dynamics?}

\author{Niels Asger Mortensen}

\address{MIC -- Department of Micro and Nanotechnology,
Nano$\bullet$DTU, Technical University of Denmark, {\O}rsteds
Plads, DTU-building 345 east, DK-2800 Kongens Lyngby, Denmark.}

\email{nam@mic.dtu.dk}

\homepage{http://www.mic.dtu.dk/nam}

\begin{abstract}
Wave chaos is a concept which has already proved its practical
usefulness in design of double-clad fibers for cladding-pumped
fiber lasers and fiber amplifiers. In general, classically chaotic
geometries will favor strong pump absorption and we address the
extent of chaotic wave dynamics in typical air-clad geometries.
While air-clad structures supporting sup-wavelength convex
air-glass interfaces (viewed from the high-index side) will
promote chaotic dynamics we find guidance of regular
whispering-gallery modes in air-clad structures resembling an
overall cylindrical symmetry. Highly symmetric air-clad structures
may thus suppress the pump-absorption efficiency $\eta$ below the
ergodic scaling law $\eta\propto A_{\rm c}/A_{\rm cl}$, where
$A_{\rm c}$ and $A_{\rm cl}$ are the areas of the rare-earth doped
core and the cladding, respectively.
\end{abstract}

\ocis{(060.2280) Fiber optics and optical communications : Fiber design and fabrication, (060.2320)  Fiber optics amplifiers and oscillators, (140.3510) Lasers and laser optics : Lasers, fiber.} 




\section{Introduction}

Cladding-pumped fibers are in general considered interesting
candidates for use as high-power fiber lasers and
amplifiers~\cite{Tunnermann:2005,Muller:2006,Canning:2006,Limpert:2006}.
Fiber lasers provide the advantages of high power levels combined
with very high beam quality. The successful scaling of power
levels for fiber lasers is powered by an ongoing technological
progress along several lines including the development of diode
lasers, pump coupling schemes, and improved double-clad fibers.

\begin{figure}[b!]
\begin{center}
\epsfig{file=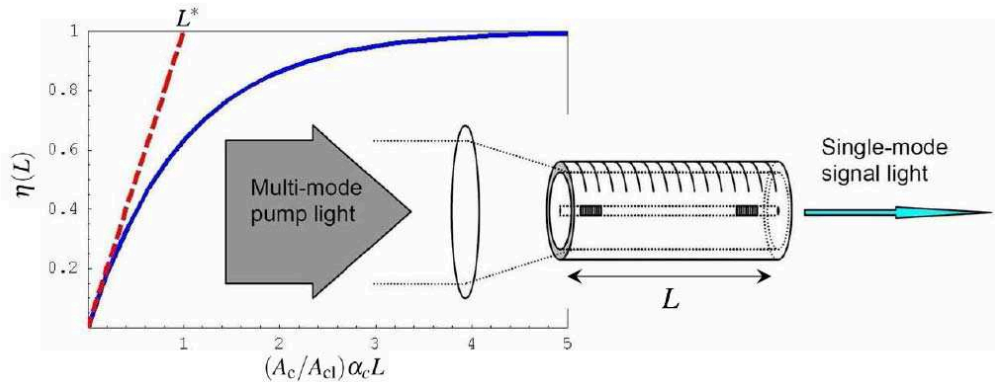, width=\textwidth,clip}
\end{center}
\caption{Plot of the pump-absorption efficiency $\eta(L)$. The
solid line shows the ergodic limit, Eq.~(\ref{eq:eta2}), while the
dashed line indicates the initial dynamics and the critical length
$L^*$, Eq.~(\ref{eq:L*2}). The inset illustrates a setup employing
a double-clad fiber for conversion of multi-mode light with a low
beam quality into single-mode light with a high beam quality.}
\label{fig1}
\end{figure}

The inset in Fig.~\ref{fig1} shows a schematic presentation of a
cladding-pumped fiber laser. The double-clad approach was
pioneered by Snitzer and co-workers in the late 1980's (see
Ref.~\cite{Snitzer:1989} and references therein) and the basic
concept of the double-clad fiber laser is to facilitate conversion
of multi-mode light into a single-mode beam of high quality. The
pump light is corralled by the double-clad and while light
propagates down the fiber, photons are absorbed and converted to
stimulated emission by a rare-earth doped fiber core at the center
of the inner cladding. Typically, the laser cavity is formed by
inscribing Bragg reflectors in the fiber core.

Here, we focus on air-clad technology~\cite{DeGiovanni:1999} which
has greatly advanced the potential of double-clad
fibers~\cite{Wadsworth:2000,Furusawa:2001,Wadsworth:2003,Limpert:2003,Bouwmans:2003,Limpert:2005}.
The large index contrast between air and silica makes the cladding
support pump light of very high numerical
aperture~\cite{Issa:2004}, thus facilitating pumping by relatively
cheap high-NA pump sources. For the corralled pump light, skew
rays~\cite{Aslund:2006}, or whispering gallery modes (WGMs), are
well-known obstacles to efficient pump absorption since they do
not have a significant overlap with the core. Such modes are
particular inherent to cylinder symmetric fiber geometries, such
as standard MCVD-fabricated fibers. The problem may be
circumvented by breaking the high cylindrical symmetry and
D-shaped cladding geometries~\cite{Doya:2001} have become a
standard approach to suppression of skew rays.

\begin{figure}[t!]
\begin{center}
\epsfig{file=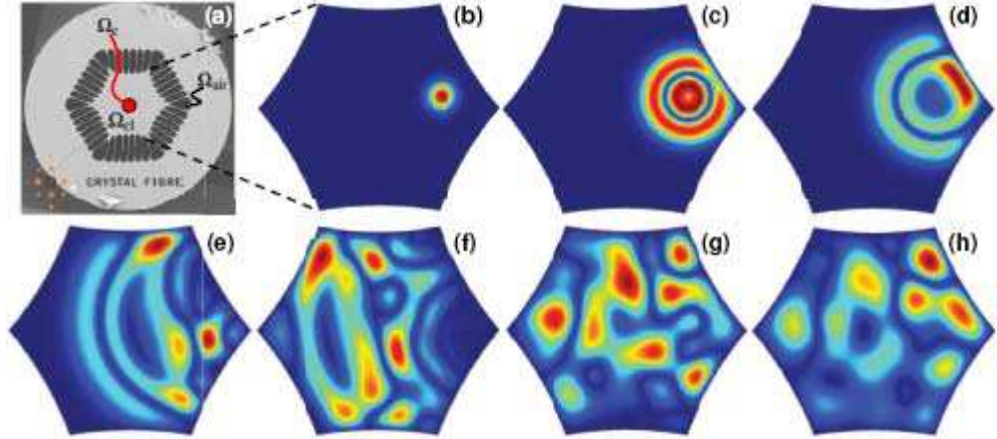, width=\textwidth,clip}
\end{center}
\caption{Panel~(a) shows a micrograph of an air-clad photonic
crystal fiber made in silica (courtesy Crystal Fibre A/S). The
pump light is corralled by an air clad consisting of 42 air holes
(dark regions) separated by sub-micron silica bridges. The
diameter of the inner cladding is of the order $125\,{\rm \mu m}$.
Panels~(b) to (h) show time-dependent finite-element simulations
of a scalar wave equation at different times (click panel to view
animation, 3.3 Mbyte) for the initial Gaussian excitation shown in
panel~(b).} \label{fig2}
\end{figure}

Pump absorption in air-clad fibers is typically very high and it
has been speculated~\cite{Broeng:2004} that this is coursed by
chaotic ray dynamics. Non-integrable air-clad geometries seem to
be inherent to the fabrication method where the shapes of the air
holes arise in an interplay and competition of e.g. the glass
viscosity and glass-air surface tension. Fig.~\ref{fig2}
illustrates one example of an air-clad fiber~\cite{Limpert:2003}
which has a classically chaotic geometry with a convex boundary
(from the point of view of the pump light) which will scatter
incident waves in a highly non-deterministic way. Time-dependent
finite-element simulations of a scalar wave equation in a
corresponding confined geometry, panels (b) to (h), do indeed
support this picture of chaotic wave dynamics. Indeed, the pump
absorption observed in Ref.~\cite{Limpert:2003} is much stronger
than for an equivalent circular geometry.

In this work we use finite-element simulations to show how the
wave dynamics has fingerprints inherited from the classically
chaotic geometry. In particular, we consider the statistics of the
mode-overlap with the core and argue that air-clad fibers tend to
be chaotic by nature in the case of sup-wavelength convex air-clad
features.

\section{Formalism}

For simplicity we consider the simplified problem with a spatially
homogeneous inner cladding region $\Omega_{\rm cl}$ with
dielectric function $\varepsilon_{\rm cl}=n_{\rm cl}^2$ and a
rare-earth doped fiber core region $\Omega_{\rm c}$ with
dielectric function $\varepsilon_{\rm c}$, see panel~(a) in
Fig.~\ref{fig2}. The inner cladding may be defined by several air
regions $\Omega_{\rm air}$. The complex-valued dielectric function
is then of the form
\begin{equation}\label{eq:epsilon}
\varepsilon(\rrr)=\varepsilon'(\rrr)+i\varepsilon''(\rrr)=\left\{\begin{array}{ccc}\varepsilon_{\rm
c}'+i\varepsilon_{\rm c}'' &,& \rrr\in \Omega_{\rm c},\\\\
\varepsilon_{\rm cl}'&,&\rrr \in \Omega_{\rm cl},\\\\
1&,&\rrr \in \Omega_{\rm air},
\end{array}\right.
\end{equation}
where for the rare-earth doped core $\varepsilon_{\rm c}''\ll
\varepsilon_{\rm c}'$ so that absorption is a weak perturbation.
In the following we consider temporal harmonic modes which are
governed by the wave equation
\begin{equation}\label{eq:wave}
\nablabf\times\nablabf\times \big|\EEE_m\big> = \varepsilon
\left(\frac{\omega_m}{c}\right)^2\big|\EEE_m\big>,\quad m=1, 2, 3,
\ldots
\end{equation}
where $\omega_m$ is the angular frequency and $c$ is the speed of
light. For a fiber geometry, translational invariance along the
fiber axis results in modes of the plane-wave form $e^{i(\beta
z-\omega t)}$ with the dispersion relation $\omega(\beta)$ found
by solving Eq.~(\ref{eq:wave}). Typically, the dispersion
properties are obtained through numerical solutions of
Eq.~(\ref{eq:wave}), but perturbation theory is another highly
valuable tool in analyzing the consequences of minor perturbations
such as variations $\Delta\varepsilon$ in the dielectric function.
For a fixed frequency, we apply standard first-order perturbation
theory to predict the corresponding shift $\Delta\beta$ in the
propagation constant $\beta$,
\begin{equation}\label{eq:Deltabeta}
\Delta\beta_m=
\left(\frac{\partial\omega_m}{\partial\beta}\right)^{-1}\frac{\omega_m}{2}
\frac{\big< \EEE_m\big|\Delta\varepsilon\big|\EEE_m\big>}{\big<
\EEE_m\big|\varepsilon\big|\EEE_m\big>}
\end{equation}
where $v_g=\partial\omega/\partial\beta$ is the group velocity
which is mathematically introduced through the use of the chain
rule~\cite{Johnson:2002}.

\section{Pump absorption}

Obviously, a complex dielectric function as in
Eq.~(\ref{eq:epsilon}) will for a fixed and real-valued frequency
result in a complex valued propagation constant
$\beta=\beta'+i\beta''$. The interpretation of the imaginary part
as a damping becomes obvious when expecting the intensity
$I(z,t)\propto \big|e^{i(\beta z-\omega t)}\big|^2 =
e^{-2\beta''z}=e^{-\alpha z}$ where we have introduced the damping
parameter $\alpha=2\beta''$. In the following we will consider the
case where the imaginary part in Eq.~(\ref{eq:epsilon}) is a small
perturbation to the real part, i.e. $\varepsilon''\ll
\varepsilon'$ so that $\beta''\ll \beta'$. From
Eq.~(\ref{eq:Deltabeta}) we immediately get
\begin{equation}\label{eq:alpha}
\alpha_m= 2\beta_m''=f_m\times\frac{\omega}{v_{g,m}}
\frac{\varepsilon_{\rm c}''}{\varepsilon_{\rm c}'}\simeq
f_m\times\alpha_{\rm c}
\end{equation}
where we have introduced the relative optical overlap with the
core
\begin{equation}\label{eq:fm}
f_m= \frac{\big< \EEE_m\big|\varepsilon\big|\EEE_m\big>_{\rm
c\quad}}{\big< \EEE_m\big|\varepsilon\big|\EEE_m\big>_{\rm c+cl}}=
\frac{\big< \EEE_m\big|\DDD_m\big>_{\rm c\quad}}{\big<
\EEE_m\big|\DDD_m\big>_{\rm c+cl}}
\end{equation}
with $\big|\DDD\big>=\varepsilon\big|\EEE\big>$ being the
displacement field. In the last equality of Eq.~(\ref{eq:alpha})
we have approximated the group velocity by that in the homogenous
core material, thus allowing for the introduction of the core
material absorption parameter $\alpha_c$. The interpretation of
Eq.~(\ref{eq:alpha}) is straight forward; the absorption of a
given mode is very intuitively given by $\alpha_c$ of the core
material weighted by the relative optical overlap $f$ of the mode
with the core. Now, since the cladding diameter by far exceeds the
wavelength of the light we will have a continuum of cladding
states, even for typical low-NA air-clad
fibers~\cite{Mortensen:2003}, so studying the characteristics of
the individual modes is cumbersome. Instead it is common to
quantify pump absorption by a single parameter, i.e. the pump
absorption efficiency as indicated in Fig.~\ref{fig1}
\begin{equation}\label{eq:eta1}
\eta(L) = \int_0^\infty d\alpha\, P(\alpha) \left[1 - \exp\left(-
\alpha L \right)\right]\simeq \int_0^\infty df\, P(f) \left[1 -
\exp\left(- f \times \alpha_c L \right)\right].
\end{equation}
Here, $P(\alpha)$ is the distribution of attenuation coefficients
while $\left[1 - \exp\left(- \alpha L \right)\right]$ is the
absorption efficiency for the individual mode for a length $L$ of
propagation. In the second equality we have used the approximation
in Eq.~(\ref{eq:alpha}) and introduced the distribution $P(f)$ of
the overlap integrals defined by Eq.~(\ref{eq:fm}). With the form
of Eq.~(\ref{eq:eta1}) we have implicitly assumed a uniform
excitation at the input facet of the fiber so that each mode has
the same initial amplitude and initially carries the same power.
Furthermore, mode mixing due to longitudinal non-uniformities,
including bending, is obviously not taken into account. In that
sense Eq.~(\ref{eq:eta1}) gives a lower bound for the pump
absorption efficiency.

A general evaluation of the integral of course requires that we
know the distribution function in detail. However, the initial
decay is easily found by expanding the square-bracket part of the
integrand. This conveniently leaves us with a result in terms of
the moments $\left<f^n\right>= \int_0^\infty df\, P(f) f^n$ of the
distribution function,
\begin{equation}\label{eq:eta2}
\eta(L) =\left<f\right> \alpha_c L- \frac{1}{2} \left<f^2\right>
\left(\alpha_c L\right)^2 + {\cal O}\big([\alpha_c L]^3\big).
\end{equation}
Obviously, the higher a mean value $\left<f\right>$ and the
smaller a spread $\left<f^2\right>$ the more efficient a pump
absorption for a given length $L$. From Eq.~(\ref{eq:eta2}) we may
now easily estimate the critical length of fiber $L^*$ required to
achieve an efficient conversion of the pump. From $\eta(L^*)\simeq
100\%$ we get
\begin{equation}\label{eq:L*1}
L^* \simeq \left<f\right>^{-1} \times \alpha_c^{-1}
\end{equation}
demonstrating how fiber geometries with a large mean value
$\left<f\right>$ will facilitate fiber lasers with a shorter laser
cavity. In the following we will study chaotic means for enhancing
the mean value and thus decreasing $L^*$.

\begin{figure}[t!]
\begin{center}
\epsfig{file=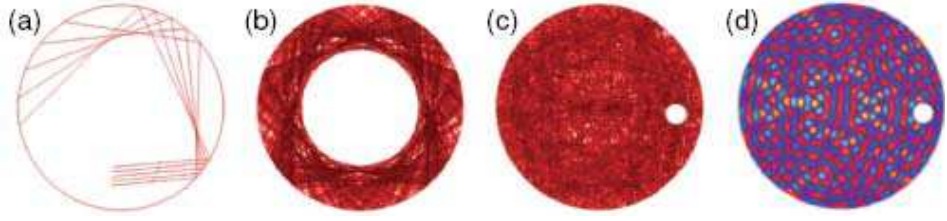, width=\textwidth,clip}
\end{center}
\caption{Ray-tracing dynamics illustrating the different response
of integrable geometries, panels (a,b), and non-integrable chaotic
geometries, panels (c,d). Panels (a) shows initial ray
trajectories in the transverse section of an integrable geometry
corresponding to an idealized multi-mode fibers having perfect
circular symmetry. Panel~(b) shows the regular whispering-gallery
like pattern that results after propagating the initial rays in
panel~(a) for a long time ($>10^3$ round trips). Panel~(c) shows
corresponding results for a non-integrable chaotic geometry
including a circular scatterer. Note that the ray pattern results
from the same initial conditions as for panel~(b). Panel~(d) shows
a corresponding 'scarred' wave function obtained from a
finite-element solution of a scalar wave equation.} \label{fig3}
\end{figure}

\section{Chaos and the ergodic hypothesis}

The term 'chaos' is used in many contexts of mathematics and
physics, but here it is used with reference to the wave dynamics
in a confined geometry. Since the double clad structures that we
are considering have dimensions much larger than the wavelength it
is quite natural to illustrate the signatures of chaos by means of
the classical ray dynamics that occurs if we consider geometrical
optics. In Fig.~\ref{fig3} we illustrate the important difference
between regular integrable geometries and their non-integrable
chaotic counterparts. The circular geometry in panel~(a) is an
excellent example where (assuming total internal reflection) we
may easily advance the rays in time, keeping in mind that the
angle of reflection equals the angle of incidence. Panel~(b) shows
the corresponding whispering-gallery like ray pattern that results
at a much later time. From a pump-absorption point of view the
problem is of course that the rays do not visit the center of the
geometry (where the absorbing core is typically located) and
changing the initial conditions (by e.g. changing the angle of
incidence) does not necessarily change this picture significantly.
Panel~(c) shows a non-integrable chaotic geometry where the
inclusion of small circular scatterer completely changes the ray
dynamics compared to panel~(b) even though the two trajectory
patterns originate from the same initial conditions. If we focus
on the wave nature of light in classically chaotic geometries then
chaos reveals itself in several ways including the 'scarred'
nature of the wave functions~\cite{Doya:2002} as illustrated by
the example in panel~(d). As for the ray dynamics in panel~(c) the
wave functions also tend to explore the entire phase space rather
than being of the whispering-gallery like nature seen in
panel~(b).

From a pump absorption point of view, chaos completely changes the
distribution $P(f)$ of overlaps. Assuming that the eigenfields
are, on average, spatially fully homogenized, then
Eq.~(\ref{eq:fm}) reduces to
\begin{equation}
f_0=\frac{\varepsilon_{\rm c}A_{\rm c}}{\varepsilon_{\rm c}A_{\rm
c}+\varepsilon_{\rm cl}A_{\rm cl}}=\frac{A_{\rm c}}{A_{\rm
cl}}+{\cal O}\big([A_{\rm c}/A_{\rm cl}]^2\big)+{\cal
O}\big(\varepsilon_{\rm c}-\varepsilon_{\rm cl}\big)
\end{equation}
where $A_{\rm c}=\int_{\Omega_{\rm c}}d\rrr$ and $A_{\rm
cl}=\int_{\Omega_{\rm cl}}d\rrr$ are the areas of the core and
cladding regions, respectively. In the following we use the
definition $f_0\equiv A_{\rm c}/A_{\rm cl}$. The above assumption
corresponds to the ergodic limit where all modes visit all parts
of the phase space equally. We then have $P(f)\simeq
\delta(f-f_0)$ so that $\big<f\big>\simeq f_0$ and due to the
simple properties of the Dirac delta function Eq.~(\ref{eq:eta1})
is now easily evaluated with the result
\begin{equation}\label{eq:eta3}
\hspace{2.6cm}\eta(L) \simeq 1 - \exp\left(- f_0 \alpha_c L
\right)\quad\quad (\textrm{ergodic limit})
\end{equation}
showing a single-exponential behavior as illustrated in
Fig.~\ref{fig1}. Finally, Eq.~(\ref{eq:L*1}) reduces to
\begin{equation}\label{eq:L*2}
L^* \simeq f_0^{-1}\times \alpha_c^{-1}\simeq\frac{A_{\rm
cl}}{A_{\rm c}} \times \alpha_c^{-1}
\end{equation}
which is the ultimate scaling which can be achieved for the
absorption of the pump.

\begin{figure}[t!]
\begin{center}
\epsfig{file=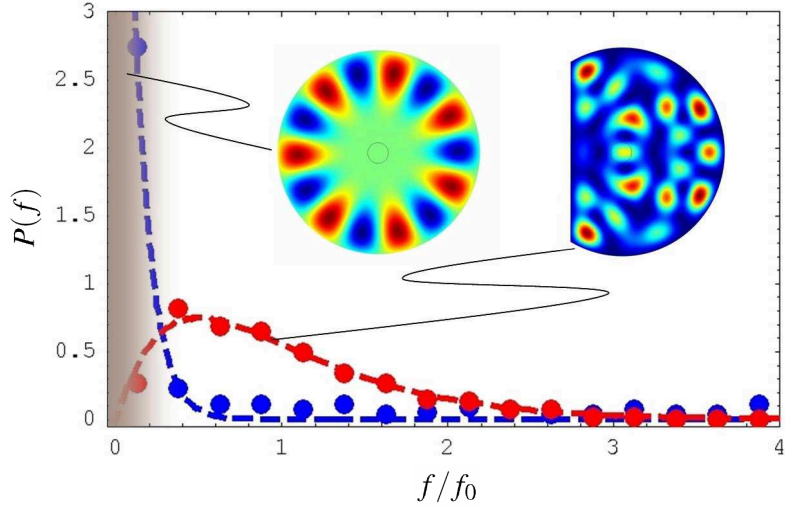, width=0.8\textwidth,clip}
\end{center}
\caption{Distribution functions for a regular integrable and
non-integrable chaotic geometry. The dashed lines shows numerical
fits to Eq.~(\ref{eq:ffit}) with $\big<f\big>/f_0\sim 9\%$ and
$\big<f\big>/f_0\sim 100\%$ for the circular and D-shape
geometries, respectively. } \label{fig4}
\end{figure}

\section{Overlap statistics}

In the following we illustrate the difference between an
integrable geometry and a chaotic geometry by comparing rods of
glass in air with a circular and D-shaped cross section,
respectively. We use a finite-element method (COMSOL Multiphysics)
to calculate fully vectorial eigenmodes of Eq.~(\ref{eq:wave})
from which we evaluate Eq.~(\ref{eq:fm}) numerically. Grouping the
data into bins of width $f_0/4$ we arrive at the histogram shown
in Fig.~\ref{fig4}. For simplicity we have considered
$\varepsilon_{\rm cl}=\varepsilon_{\rm c}=(1.45)^2$ and for the
wavelength we have used $\lambda/R=0.1$ with $R$ being the radius
of the cladding. Furthermore, the core is placed in the center of
the fiber and $A_{\rm c}/A_{\rm cl}=0.01$. Results for the
circular case are shown in blue while the red results are for a
corresponding D-shaped geometry. The dashed lines are numerical
fits to the following normalized distribution functions
\begin{equation}\label{eq:ffit}
P_{\rm fit}(f)=\left\{\begin{array}{ccccc}
\big<f\big>^{-1}\exp\left(-f/\big<f\big>\right)&,&\big<f\big>/f_0\sim 9\% &,&(\textrm{circular}) ,\\\\
4f\big<f\big>^{-2}\exp\left(-2f/\big<f\big>\right)&,&
\big<f\big>/f_0\sim 100\%  &,& (\textrm{D-shape}) ,
\end{array}\right.
\end{equation}
with the mean value $\big<f\big>$ used as fitting parameter.
Fig.~\ref{fig4} is illustrating the close to exponential
distribution in the case of the integrable geometry while for the
chaotic geometry the weight is shifted to much larger values with
a suppression of modes with a poor overlap. The insets show
typical modes for the two geometries. For the circular geometry
the peak in the distribution originates from WPGs which are
suppressed in the D-shaped geometry where modes are of the
'scarred' nature with an average overlap $\big<f\big>\sim f_0$ in
agreement with the ergodic hypothesis. Obviously, in the context
of $\eta$ the chaotic structure is superior to the integrable one
as also emphasized previously in the
literature~\cite{Doya:2001,Kouznetsov:2001,Kouznetsov:2002,Kouznetsov:2002a}.

\begin{figure}[t!]
\begin{center}
\epsfig{file=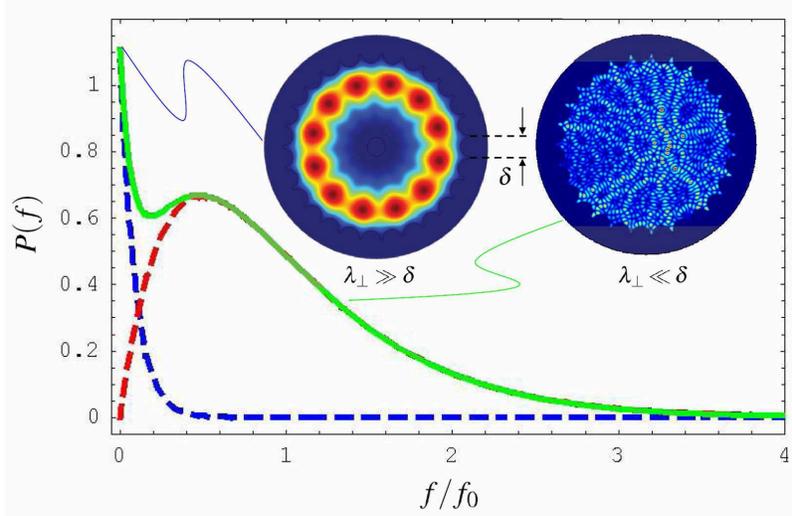, width=0.8\textwidth,clip}
\end{center}
\caption{Qualitative illustration of the distribution function for
an air-clad structure displaying partly integrable
($\lambda_\perp\gg \delta$) and partly chaotic dynamics
($\lambda_\perp\ll \delta$). The insets illustrate the two classes
of modes representing a vanishing overlap $f\sim 0$ and a finite
overlap $f\sim f_0$, respectively. } \label{fig5}
\end{figure}

\begin{figure}[t!]
\begin{center}
\epsfig{file=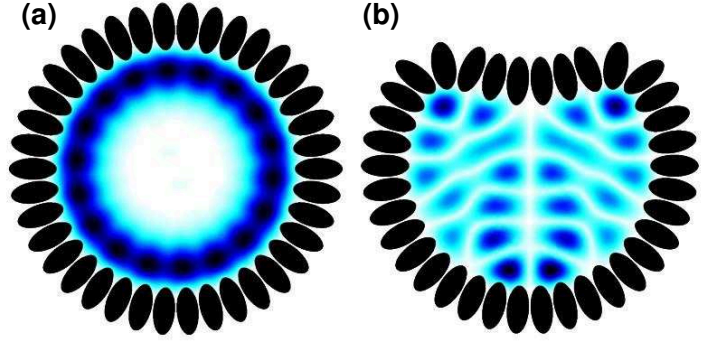, width=0.7\textwidth,clip}
\end{center}
\caption{Panel~(a) shows a whispering-gallery mode in a
ring-shaped air-clad geometry while panel~(b) shows a
corresponding mode for a D-shaped air-clad geometry.} \label{fig6}
\end{figure}

A quite natural question now arises; are air clad structures truly
chaotic? As we shall see the answer to this inquiry depends a lot
on the geometry into which the air holes are arranged. For the
simulations in Fig.~\ref{fig2} we neglected the sub-micron silica
bridges separating the sup-micron convex air-clad features. At
first this appears reasonable if the bridges are sub-wavelength.
However, one should not compare to the free-space wavelength
$\lambda$ (or $\lambda/n_{\rm cl}$ for that sake), but one should
rather consider the effective transverse wavelength
$\lambda_\perp$ which may exceed $\lambda$ significantly.
Projecting the problem onto the transverse cross section of the
fiber we get
\begin{equation}
\lambda_{\perp}\simeq \frac{\lambda}{\sqrt{n_{\rm cl}^2-n_{{\rm
eff}}^2}}
\end{equation}
where $n_{{\rm eff}} = c\beta/\omega_m$ is the effective index of
mode $m$. Now, let $\delta$ be a characteristic length scale for
the air-clad features, such as the bridge width or the air-hole
diameter. The modes may now be classified by comparing
$\lambda_\perp$ with $\delta$;
\begin{itemize}
\item For $\lambda_\perp\gg \delta$ the electromagnetic
field effectively experiences a smooth boundary defined by the air
clad and the wave dynamics may be {\bf integrable} or {\bf
chaotic} depending on the effective shape of the corral.
\item For $\lambda_\perp\ll \delta$ the field explores all the
detailed features and in particular the convex air-clad boundary
so that the wave dynamics turn {\bf chaotic}.
\end{itemize}
While the geometry in Fig.~\ref{fig2} is chaotic in both limits
one can easily imagine more regular air-clad structures which will
only be 'partly' chaotic, i.e. low-NA modes will experience an
integrable confinement geometry while high-NA modes will
experience a chaotic geometry. Indeed, the insets in
Fig.~\ref{fig5} illustrate this perfectly for a geometry very much
resembling the over-all cylindrical symmetry of the air-clad
structure employed in more recent works such as
Ref.~\cite{Limpert:2006a}. The air-clad structure consists of 27
air-holes distributed evenly in a cylindrical symmetry with radius
$R$. The modes are obtained by fully-vectorial finite-element
solutions of Eq.~(\ref{eq:wave}) for a wavelength of
$\lambda/R=0.01$. While a quantitative numerical simulation of the
full statistics and the distribution function is cumbersome we
imagine that the two classes of modes will give rise to a
distribution function qualitatively interpolating the two limits
in Eq.~(\ref{eq:ffit}) as indicated schematically in
Fig.\ref{fig5}. The existence of skew rays and WPGs, represented
by the exponential peak near $f\sim 0$, will eventually suppress
the average value $\big<f\big>$ below the ergodic limit given by
$f_0$ thus compromising the pump-absorption efficiency $\eta$ and
the critical absorption length $L^*$. Deforming the ring-shaped
air-clad slightly, or introducing an additional air hole as in
panel~(c) or (d) of Fig.~\ref{fig3}, would suppress the
exponential peak at $f\sim 0$ and restore the distribution
function with $\big<f\big>\sim f_0$ characteristic of chaotic wave
dynamics. Fig.~\ref{fig6} illustrates how WPGs may be suppressed
by constructing the air-clad in a D-shaped fashion.

\section{Conclusion}

We have discussed wave chaos in the context of the pump-absorption
efficiency $\eta$ in air-clad fibers which are now being widely
used for high-power cladding-pumped fiber lasers. Starting from
the wave equation we rigorously derive a result for $\eta$ which
in the ergodic limit scales as $\eta\propto A_{\rm c}/A_{\rm cl}$,
where $A_{\rm c}$ and $A_{\rm cl}$ are the areas of the rare-earth
doped core and the cladding, respectively. While air-clad
structures supporting sup-wavelength convex air-clad features will
promote chaotic dynamics we also show how this may be jeopardized
by whispering-gallery modes in air-clad structures resembling an
overall cylindrical symmetry. In the absence of mode-mixing
mechanisms, highly symmetric air-clad structures may thus suppress
the pump-absorption efficiency $\eta$ below the ergodic scaling
law.

\section*{Acknowledgements}

Publication of this work is financially supported by the
\emph{Danish Council for Strategic Research} through the
\emph{Strategic Program for Young Researchers} (grant no:
2117-05-0037). J. Broeng, A. Petersson, M.~D. Nielsen, J. Riis
Folkenberg, P.~M.~W. Skovgaard, C. Jakobsen, H. Simonsen, and J.
Limpert are acknowledged for sharing their insight in experiments,
theory, design, and fabrication of air clad fibers.


\begin{thebibliography}{10}
\newcommand{\enquote}[1]{``#1''}
\expandafter\ifx\csname url\endcsname\relax
  \def\url#1{\texttt{#1}}\fi
\expandafter\ifx\csname
urlprefix\endcsname\relax\def\urlprefix{URL }\fi
\providecommand{\eprint}[2][]{\url{#2}}

\bibitem{Tunnermann:2005}
A.~T{\"u}nnermann, T.~Schreiber, F.~R{\"o}ser, A.~Liem,
S.~H{\"o}fer,
  H.~Zellmer, S.~Nolte, and J.~Limpert, \enquote{The renaissance and bright
  future of fibre lasers,} J. Phys. B-At. Mol. Opt. Phys. \textbf{38}(9), S681
  -- S693 (2005).

\bibitem{Muller:2006}
H.-R. M{\"u}ller, J.~Kirchhof, V.~Reichel, and S.~Unger,
\enquote{Fibers for
  high-power lasers and amplifiers,} C. R. Phys. \textbf{7}(2), 154 -- 162
  (2006).

\bibitem{Canning:2006}
J.~Canning, \enquote{Fibre lasers and related technologies,} Opt.
Lasers Eng.
  \textbf{44}(7), 647 -- 676 (2006).

\bibitem{Limpert:2006}
J.~Limpert, F.~R{\"o}ser, T.~Schreiber, I.~Manek-H{\"o}nninger,
F.~Salin, and
  A.~T{\"u}nnermann, \enquote{Ultrafast high power fiber laser systems,} C. R.
  Phys. \textbf{7}(2), 187 -- 197 (2006).

\bibitem{Snitzer:1989}
E.~Snitzer, \enquote{Rare-earth fiber lasers,} J. Less-Common
Metals
  \textbf{148}(1-2), 45 -- 58 (1989).

\bibitem{DeGiovanni:1999}
D.~DiGiovanni, \enquote{Method of making a cladding pumped fiber
structure,} US
  Patent \# 5873923  (1999).

\bibitem{Wadsworth:2000}
W.~J. Wadsworth, J.~C. Knight, W.~H. Reeves, P.~S.~J. Russell, and
J.~Arriaga,
  \enquote{Yb3+-doped photonic crystal fibre laser,} Electron. Lett.
  \textbf{36}(17), 1452 -- 1454 (2000).

\bibitem{Furusawa:2001}
K.~Furusawa, A.~Malinowski, J.~H.~V. Price, T.~M. Monro, J.~K.
Sahu,
  J.~Nilsson, and D.~J. Richardson, \enquote{Cladding pumped Ytterbium-doped
  fiber laser with holey inner and outer cladding,} Opt. Express
  \textbf{9}(13), 714 -- 720 (2001).

\bibitem{Wadsworth:2003}
W.~J. Wadsworth, R.~M. Percival, G.~Bouwmans, J.~C. Knight, and
P.~S.~J.
  Russel, \enquote{High power air-clad photonic crystal fibre laser,} Opt.
  Express \textbf{11}(1), 48 -- 53 (2003).

\bibitem{Limpert:2003}
J.~Limpert, T.~Schreiber, S.~Nolte, H.~Zellmer, A.~T{\"u}nnermann,
R.~Iliew,
  F.~Lederer, J.~Broeng, G.~Vienne, A.~Petersson, and C.~Jakobsen,
  \enquote{High-power air-clad large-mode-area photonic crystal fiber laser,}
  Opt. Express \textbf{11}(7), 818 -- 823 (2003).

\bibitem{Bouwmans:2003}
G.~Bouwmans, R.~M. Percival, W.~J. Wadsworth, J.~C. Knight, and
P.~S.~J.
  Russell, \enquote{High-power Er : Yb fiber laser with very high numerical
  aperture pump-cladding waveguide,} Appl. Phys. Lett. \textbf{83}(5), 817 --
  818 (2003).

\bibitem{Limpert:2005}
J.~Limpert, N.~D. Robin, I.~Manek-H{\"o}nninger, F.~Salin,
F.~R{\"o}ser, A.~Liem,
  T.~Schreiber, S.~Nolte, H.~Zellmer, A.~T{\"u}nnermann, J.~Broeng,
  A.~Petersson, and C.~Jakobsen, \enquote{High-power rod-type photonic crystal
  fiber laser,} Opt. Express \textbf{13}(4), 1055 -- 1058 (2005).

\bibitem{Issa:2004}
N.~A. Issa, \enquote{High numerical aperture in multimode
microstructured
  optical fibers,} Appl. Optics \textbf{43}(33), 6191 -- 6197 (2004).

\bibitem{Aslund:2006}
M.~Aslund, S.~D. Jackson, J.~Canning, A.~Teixeira, and
K.~Lyytikainen-Digweed,
  \enquote{The influence of skew rays on angular losses in air-clad fibres,}
  Opt. Commun. \textbf{262}(1), 77 -- 81 (2006).

\bibitem{Doya:2001}
V.~Doya, O.~Legrand, and F.~Mortessagne, \enquote{Optimized
absorption in a
  chaotic double-clad fiber amplifier,} Opt. Lett. \textbf{26}(12), 872 -- 874
  (2001).

\bibitem{Broeng:2004}
J.~Broeng, G.~Vienne, A.~Petersson, P.~M.~W. Skovgaard, J.~R.
Folkenberg, M.~D.
  Nielsen, C.~Jakobsen, H.~Simonsen, and N.~A. Mortensen, \enquote{Air-clad
  photonic crystal fibers for high-power single-mode lasers,} Proc. SPIE
  \textbf{5335}, 192 -- 201 (2004).

\bibitem{Johnson:2002}
S.~G. Johnson, M.~Ibanescu, M.~A. Skorobogatiy, O.~Weisberg, J.~D.
  Joannopoulos, and Y.~Fink, \enquote{Perturbation theory for
  \uppercase{M}axwell's equations with shifting material boundaries,} Phys.
  Rev. E \textbf{65}(6), 066611 (2002).

\bibitem{Mortensen:2003}
N.~A. Mortensen, M.~Stach, J.~Broeng, A.~Petersson, H.~R.
Simonsen, and
  R.~Michalzik, \enquote{Multi-mode photonic crystal fibers for
  \uppercase{VCSEL} based data transmission,} Opt. Express \textbf{11}(17),
  1953 -- 1959 (2003).

\bibitem{Doya:2002}
V.~Doya, O.~Legrand, F.~Mortessagne, and C.~Miniatura,
\enquote{Speckle
  statistics in a chaotic multimode fiber,} Phys. Rev. E \textbf{65}(5),
  056223 (2002).

\bibitem{Kouznetsov:2001}
D.~Kouznetsov, J.~V. Moloney, and E.~M. Wright,
\enquote{Efficiency of pump
  absorption in double-clad fiber amplifiers. I. Fiber with circular symmetry,}
  J. Opt. Soc. Am. B \textbf{18}(6), 743 -- 749 (2001).

\bibitem{Kouznetsov:2002}
D.~Kouznetsov and J.~V. Moloney, \enquote{Efficiency of pump
absorption in
  double-clad fiber amplifiers. II. Broken circular symmetry,} J. Opt. Soc. Am.
  B \textbf{19}(6), 1259 -- 1263 (2002).

\bibitem{Kouznetsov:2002a}
D.~Kouznetsov and J.~V. Moloney, \enquote{Efficiency of pump
absorption in
  double-clad fiber amplifiers. III. Calculation of modes,} J. Opt. Soc. Am. B
  \textbf{19}(6), 1304 -- 1309 (2002).

\bibitem{Limpert:2006a}
J.~Limpert, O.~Schmidt, J.~Rothhardt, F.~R{\"o}ser, T.~Schreiber,
  A.~T{\"u}nnermann, S.~Ermeneux, P.~Yvernault, and F.~Salin, \enquote{Extended
  single-mode photonic crystal fiber lasers,} Opt. Express \textbf{14}(7), 2715
  -- 2720 (2006).

\end{thebibliography}
\end{document}